\documentclass[twocolumn,english,aps,arxiv,superscriptaddress]{revtex4-1}
\usepackage{grffile}
\usepackage[latin9]{inputenc}
\usepackage{color}
\usepackage{babel}
\usepackage{graphicx}
\usepackage{epstopdf}
\PassOptionsToPackage{normalem}{ulem}
\usepackage{ulem}
\usepackage[unicode=true, bookmarks=false, breaklinks=false,pdfborder={0 0 1},colorlinks=false] {hyperref}
\usepackage{breakurl}

\begin{document}


\title{Non-trivial topology in a layered Dirac nodal-line semimetal candidate SrZnSb$_2$ with distorted Sb square nets}

\author{Jinyu Liu}
\email{Corresponding author: liujy@physics.ucla.edu}
\affiliation{Department of Physics and Astronomy, University of California, Los Angeles, California 90095, USA}
\author{Pengfei Liu}
\affiliation{Shenzhen Institute for Quantum Science and Technology and Department of Physics, Southern University of Science and Technology, Shenzhen, 518055, China}
\author{Kyle Gordon}
\affiliation{Department of Physics, University of Colorado, Boulder, CO 80309, USA}
\author{Eve Emmanouilidou}
\affiliation{Department of Physics and Astronomy, University of California, Los Angeles, California 90095, USA}
\author{Jie Xing}
\affiliation{Department of Physics and Astronomy, University of California, Los Angeles, California 90095, USA}
\author{David Graf}
\affiliation{National High Magnetic Field Laboratory, 1800 E. Paul Dirac Drive, Tallahassee, FL 32310, USA}
\author{Bryan C. Chakoumakos}
\affiliation{Quantum Condensed Matter Division, Oak Ridge National Laboratory, Oak Ridge, Tennessee 37831, USA}
\author{Yan Wu}
\affiliation{Quantum Condensed Matter Division, Oak Ridge National Laboratory, Oak Ridge, Tennessee 37831, USA}
\author{Huibo cao}
\affiliation{Quantum Condensed Matter Division, Oak Ridge National Laboratory, Oak Ridge, Tennessee 37831, USA}
\author{Dan Dessau}
\affiliation{Department of Physics, University of Colorado, Boulder, CO 80309, USA}
\author{Qihang Liu}
\affiliation{Shenzhen Institute for Quantum Science and Technology and Department of Physics, Southern University of Science and Technology, Shenzhen, 518055, China}
\affiliation{Center for Quantum Computing, Peng Cheng Laboratory, Shenzhen, 518055, China}
\author{Ni Ni}
\email{Corresponding author: nini@physics.ucla.edu}
\affiliation{Department of Physics and Astronomy, University of California, Los Angeles, California 90095, USA}

\begin{abstract}
Dirac states hosted by Sb/Bi square nets are known to exist in the layered antiferromagnetic AMnX$_{2}$ (A = CaSr/Ba/Eu/Yb, X=Sb/Bi) material family with the space group to be $P4/nmm$ or $I4/mmm$. In this paper, we present a comprehensive study of quantum transport behaviors, angle-resolved photoemission spectroscopy (ARPES) and first-principles calculations on SrZnSb$_2$, a nonmagnetic analogue to AMnX$_{2}$, which crystalizes in the $pnma$ space group with distorted square nets. From the quantum oscillation measurements up to 35 T, three major frequencies including $F_1 =$ 103 T, $F_2 =$ 127 T and $F_3 =$ 160 T, are identified. The effective masses of the quasiparticles associated with these frequencies are extracted, namely, $m_{1}^\ast =$ 0.1 m$_{e}$, $m_{2}^\ast =$ 0.1 m$_{e}$ and $m_{3}^\ast =$ 0.09 m$_{e}$, where m$_{e}$ is the free electron mass. From the three band Lifshitz-Kosevich fit, the Berry phases accumulated along the cyclotron orbit of the quasiparticles are $0.06\pi$, $1.2\pi$ and $0.74\pi$ for $F_{1}$, $F_{2}$ and $F_{3}$, respectively. Combined with the ARPES data and the first-principles calculations, we reveal that $F_{2}$ and $F_{3}$ are associated with the two nontrivial Fermi pockets at the Brillouin zone edge while $F_{1}$ is associated with the trivial Fermi pocket at the zone center. In addition, the first-principles calculations further suggest the existence of Dirac nodal line in the band structure of SrZnSb$_2$.
\end{abstract}

\date{\today}
\maketitle

\section{Introduction}
Topological semimetals including Dirac semimetals, Weyl semimetals and topological node-line semimetals, representing new quantum states of matter, have attracted enormous attention in condensed matter physics \cite{TI_review, Na3Bi_predict, Cd3As2_predict, Na3Bi_ARPES, Cd3As2_ARPES, Cd3As2_ARPES2, Cd3As2_ARPES3, Cd3As2_MR, Weyl_predict, Weyl_ARPES1, Weyl_ARPES2, Node-line_predict, ZrSiS_Node-line_ARPES, PbTaSe2, PtSn4, ZrSiX, CaAgX, CaTX, Weyl-II, MoTe2_Weyl, WTe2_Weyl2, WTe2_Weyl}. In these systems, the low energy excitation of quasiparticles are of relativistic nature, i.e. they obey Dirac or Weyl equations \cite{TS_review}. From the band structure point of view, there exist linear bands with nontrivial topology near the Fermi level in the bulk state of these materials. Materials with such nontrivial topological bands may be found if guided by certain chemistry principles \cite{TQ_chemistry, chemistry_principle}.

Topological bands generated by square nets of main group elements have been reported in many layered materials such as AMnX$_{2}$ (A = CaSr/Ba/Eu/Yb, X=Sb/Bi) \cite{Park_SrMnBi2, Wang_CaMnBi2, Borisenko_YbMnBi2, Masuda_EuMnBi2, Li_BaMnBi2, Wang_BaMnBi2, Liu_BaMnSb2, Liu_SrMnSb2, He_CaMnSb2, Huang_BaMnSb2, Wang_YbMnSb2, Kealhofer_YbMnSb2} and the WHM-type (W=Zr/Hf/La, H=Si/Ge/Sn/Sb, M=S/Se/Te) materials \cite{Node-line_predict, ZrSiS_Node-line_ARPES, ZrSiX, dHvA_ZrGeM, Schoop_CeSbTe}. The AMnX$_{2}$ materials feature a common layered structure with alternative stacking of MnX$_4$ tetrahedral layers and A-X-A spacing layers. In the MnX$_4$ layers, Mn atoms sit at the centers of edge-sharing tetrahedrons, and develop antiferromagnetic (AFM) order with a Néel temperature around room temperature \cite{Magnetism_AMnBi2, Liu_BaMnSb2, Liu_SrMnSb2}. In the A-X-A layers, the X square net planes sandwiched by A atoms can host anisotropic Dirac or Weyl cones. The electronic structure in these materials is of quasi-2D nature because of the layered structure character. In addition, the magnetism can have an effect on the transport of relativistic fermions. In EuMnBi$_{2}$, the Dirac fermions in 2D Bi layers are confined by antiferromagnetically ordered Eu and Mn moments and give rise to bulk quantum Hall effect \cite{Masuda_EuMnBi2}. In Sr$_{1-y}$Mn$_{1-z}$Sb$_{2}$ ($y, z < 0.1$), where the Sb square net layers are distorted into Sb zig-zag chains, the AFM ordered Mn moments are canted with a net ferromagnetic (FM) component whose magnitude varies with sample, and such a FM component affects the transport properties significantly \cite{Liu_SrMnSb2}. In YbMnBi$_{2}$, the similar canted AFM order of Mn moments has been suggested to break the time reversal symmetry and lead to the emergence of type-II Weyl states \cite{Borisenko_YbMnBi2, Chinotti_YbMnBi2, Liu_YbMnBi2}.

However, it is still unclear if the antiferromagnetism plays an indispensable role for the presence of the Dirac fermion transport in AMnX$_2$ family\cite{Interplay_AMnBi2, Wang_BaZnBi2, Ren_BaZnBi2, Zhao_BaZnBi2}. The elucidation of the topology of their non-magnetic analogs will shed light on it. So far, the square nets of main group elements hosting Dirac states, which are found in the materials with space group of $I4/mmm$ or $P4/nmm$, are all undistorted except the Sb zig-zag-chain layer in Sr$_{1-y}$Mn$_{1-z}$Sb$_2$. It has been pointed out by Bradlyn, \emph{et. al.} in Ref. \cite{TQ_chemistry} that the topological behavior is insensitive to the in-plane distortion. More than 50 candidates composed by distorted square nets in space group $pnma$, including SrZnSb$_{2}$, are predicted to be topological nontrivial phases \cite{TQ_chemistry}. In this paper, we focus on SrZnSb$_2$, whose crystal structure is shown in Fig. 1(a). A previous study on SrZnSb$_{2}$ revealed linear magnetoresistance up to 9 T, suggesting the possible existence of Dirac fermion transport in the system \cite{Wang_SrZnSb2}. We have grown SrZnSb$_2$ single crystals and confirmed its orthorhombic structure with space group of $Pnma$ at both 300 K and 105 K through single crystal X-Ray diffraction\cite{SrZnSb2_prep}. We present the results from quantum transport measurements up to 35 T on SrZnSb$_{2}$ and reveal its detailed electronic structure with the input from first-principles calculations and ARPES experiments. From the quantum oscillations, we can derive three fundamental frequencies, i.e. $F_1 =$ 103 T, $F_2 =$ 127 T and $F_3 =$ 160 T and attribute them to a trivial quasi-2D Fermi pocket $\gamma$ and two nontrivial quasi-2D Fermi pockets $\alpha$ and $\beta$, respectively. The Berry phases for the three frequencies agree well with the calculated values for $\alpha$, $\beta$ and $\gamma$ pockets. In addition, our first-principles calculations suggest this compound is a nodal-line Dirac semimetal candidate, which is of high interest for the future study on this material.

\section{Experimental Methods}
Single crystals of SrZnSb$_{2}$ were grown by the flux method. Sr and Zn chunks and Sb powder were mixed at a ratio of $1:1:5$ in an alumina crucible and sealed in a quartz tube under 1/3 atmosphere of Ar gas. The tube was heated at 1050 ºC for 24 hours and slowly cooled to 650 ºC at a cooling rate of 3 ºC/hour. The excess Sb flux was then decanted by centrifuging. As shown in the inset of Fig. 1(c), shining plate-like single crystals were obtained.

To determine whether the crystal remains in the \emph{pnma} space group or undergoes some subtle structural distortion which has been observed in some of the 112 family of compounds, such as REAgSb$_2$ \cite{ReAgSb_2, LaAgSb_2}, single crystal diffraction measurement was performed. Single-crystal diffraction data were collected at 300 K and 105 K using a Rigaku XtaLAB PRO diffractometer with graphite monochromated Mo $K\alpha$ radiation ($ \lambda =$ 0.71073 {\AA}) equipped with a Dectris Pilatus 200 K detector and an Oxford N-HeliX cryocooler. Numerous crystals were screened, and full data sets were collected for three pieces. Peak indexing and integration were done using the Rigaku Oxford Diffraction CrysAlisPro software \cite{X-ray_Rigaku}. An empirical absorption correction was applied using the SCALE3 ABSPACK algorithm as implemented CrysAlisPro. WinGX and SHELXL-2013 software packages were used for data processing and structure solution and refinement \cite{X-ray_Refinement}. Crystal structure projections were made with VESTA \cite{vesta}.

The magnetotransport properties were measured in a Quantum Design Physical Property Measurement System (QD PPMS) up to 9 T, and at the National High Magnetic Field Lab (NHMFL) in Tallahassee, FL, up to 35 T. The $\rho_{xx}$ was measured by a standard four probe method. To obtain $\rho_{zz}$, the current contacts were made into a ring shape at the top and bottom of the sample surfaces ($bc$ plane) and the voltage contacts were placed at the center of the rings. The torque measurements were done using a piezoresistive cantilever connecting with a homemade Wheatstone bridge at NHMFL.

We applied density functional theory (DFT) by using the projector-augmented wave (PAW) pseudopotentials \cite{Calculation_method1} with the exchange-correlation of Perdew-Burke-Ernzerhof (PBE) form \cite{Calculation_method2} as implemented in the Vienna ab-initio Simulation Package (VASP) \cite{Calculation_method3}. The energy cutoff was chosen 1.5 times as large as the values recommended in relevant pseudopotentials. The $k$-points-resolved value of Brillouin zone (BZ) sampling was 0.02$\times$2$\pi /{\AA}$. The total energy minimization was performed with a tolerance of 10$^{-6}$ eV. The crystal structure was fully relaxed until the atomic force on each atom was less than 10$^{-2}$ eV$/{\AA}$. For the calculations with strain, atoms were relaxed with fixed lattice constants. Spin-orbit coupling (SOC) was included self-consistently throughout the calculations.

Angle-resolved photoemission spectroscopy (ARPES) measurements were conducted at the Advanced Light Source (ALS), BL4.0.3, with horizontally polarized, 98 eV light. The sample was held at 20 K during the measurements. We estimated 44 meV energy resolution from the Fermi edge of the sample. The sample cleave appeared smooth under an optical microscope.

\section{Experimental Results }
\begin{figure*}
  \centering
  \includegraphics[width=7in]{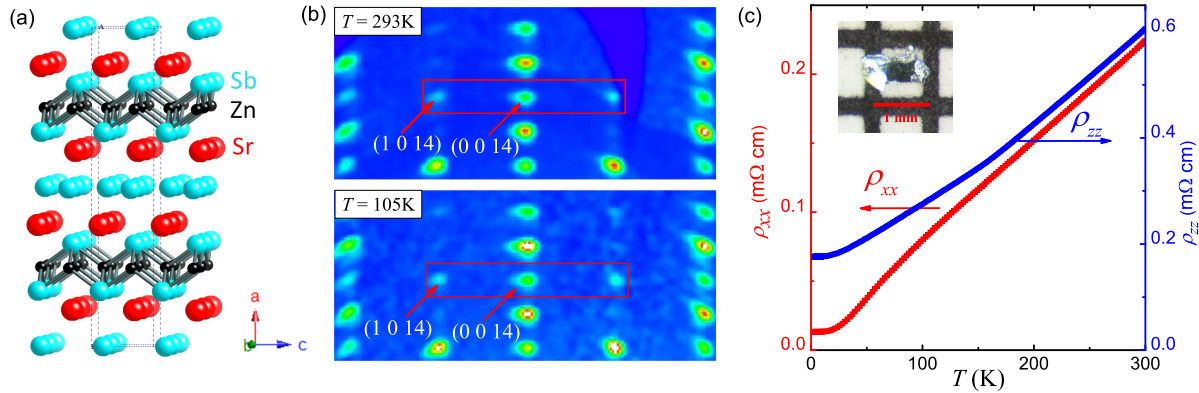}
  \caption{(a) Crystal structure of SrZnSb$_{2}$. (b) Reciprocal lattice plane views at 293 K (Top) and 105 K (Bottom). (c) In-plane resistivity $\rho_{xx}$ (red) and out-of-plane resistivity $\rho_{zz}$ (blue) as a function of temperature from 2 K to 300 K. Inset is an optical image of a piece of SrZnSb$_{2}$ single crystal placed on a millimeter grid paper.
  }
  \label{fig:Fig1}
\end{figure*}

\newcommand{\tabincell}[2]{\begin{tabular}{@{}#1@{}}#2\end{tabular}}
\begin{table}
\caption{Rietveld refinement results of SrZnSb$_2$ at 105 K and 293 K \cite{Fullprof}. Crystal size is $150\times50\times10$ $\mu$m$^3$. }
\begin{tabular}{c c c}
\hline
\hline
Formula & SrZnSb$_2$ & SrZnSb$_2$ \\
\hline
$T$ (K) &105(1)  &293(1)  \\
Crystal system & orthorhombic & orthorhombic\\
Space group & \emph{pnma} & \emph{pnma} \\
\emph{a} (\AA) &22.835(9) & 22.980(9) \\
\emph{b} (\AA) &4.3534(16) & 4.3910(13)\\
\emph{c} (\AA) &4.4009(16) & 4.4179(15)\\
\emph{V} (\AA $^{3}$) & 437.5(3) & 445.8(3)\\
\tabincell{c}{$\theta$ } &3.57-36.12 &4.43-24.71\\
\tabincell{c}{No. reflections collected } &2879 &4436\\
No. of variables &26  &26\\
$R_1$ &0.148 &0.148\\
$\omega R_2$ &0.358 &0.428\\
Goodness of fit &2.53 &3.57\\
\tabincell{c}{Largest diff. peak \\and hole (e\AA$^{-3}$)} &7.68, -6.54 &9.80, -11.66\\
\hline
\hline
\end{tabular}
\label{tab.1}
\end{table}

\begin{table}
\caption{Atomic coordinates and equivalent isotropic displacement parameters of SrZnSb$_2$ at 105(1) K and 293(1) K}
\begin{tabular}{c c c c c c}
\hline
\hline
\multicolumn{6}{c}{105 K}\\
\hline
Atom & Site & \emph{x} & \emph{y} & \emph{z} & $U_{eq}$ (\AA$^{2})$ \\[1ex]
\hline
Sr & 4c & 0.1154(2) &1/4 &0.721(1) & 0.015(1)\\
Zn & 4c & 0.2502(8) &1/4 &0.226(1) & 0.019(2)\\
Sb1 & 4c & 0.0004(4) &1/4 &0.2213(9) & 0.015(1)\\
Sb2 & 4c & 0.3231(1) &1/4 &0.7244(9) & 0.011(1)\\
\hline
\multicolumn{6}{c}{293 K}\\
\hline
Atom & Site & \emph{x} & \emph{y} & \emph{z} & $U_{eq}$ (\AA$^{2})$ \\[1ex]
\hline
Sr & 4c & 0.11514(2) &1/4 &0.723(1) & 0.023(2)\\
Zn & 4c & 0.250(1) &1/4 &0.226(1) & 0.030(3)\\
Sb1 & 4c & -0.0011(5) &1/4 &0.222(1) & 0.022(1)\\
Sb2 & 4c & 0.3232(1) &1/4 &0.720(1) & 0.014(1)\\
\hline
\hline
\end{tabular}
\label{tab.2}
\end{table}

Like many other layered materials, the overall crystal quality of the samples was imperfect as evidenced by the lack of good agreement between symmetry equivalent reflections and this disagreement could not be fully explained by twinning. Nevertheless, the best refinements (Table I) as well as reciprocal lattice plane views (Fig. 1(b)) are unable to discern any symmetry lowering between 105 K and 293 K. Neither additional reflections nor superlattice peaks were observed at 105 K. We also carefully checked the intensities of Bragg peaks at (0 0 14) and (1 0 14)/(0 1 14) at 105 K and 293 K collected on three crystals, which correspond to the ones at (0 0 7) and (1 0 7) in LaAgSb$_2$ \cite{LaAgSb_2}, respectively. In LaAgSb$_2$, charge density wave was claimed by the observation of superlattice peaks at (1 0 7). This is not the case for SrZnSb$_2$ since the ratio of (1 0 14) or (0 1 14) to (0 0 14) shows no discernable changes at 105 K and 293 K. Therefore, our single crystal X-ray results do not suggest structural distortions in SrZnSb$_2$.

The resistivity has been measured along the in-plane direction ($\rho_{xx}$) and out-of-plane direction ($\rho_{zz}$), as shown in Fig. 1(c). Both $\rho_{xx}$ and $\rho_{zz}$ show metallic behavior in the temperature range from 300 K to 2 K. The resistivity anisotropy $\rho_{xx}$/$\rho_{zz}$ increases from $\sim$3 at the room temperature to 14 at 2 K, slightly higher than 12 for CaMnBi$_2$, \cite{Awang_CaMnBi2} indicating the electronic structure is more anisotropic. Although a subtle slope change can be seen in $\rho_{zz}$ around 150 K, as we have discussed, no clear evidence of possible structural distortion across this temperature is observed from our single crystal X-ray diffraction data. For $\rho_{xx}$, the residual resistivity ratio ($\rho_{xx}$(300K)/$\rho_{xx}$(2K)) is $\sim$18, comparable to 15 for SrMnBi$_{2}$, \cite{Park_SrMnBi2} and much higher than 2 for SrZnBi$_{2}$ and 7 for BaZnBi$_{2}$, \cite{Wang_BaZnBi2, Zhao_BaZnBi2} which suggests good crystalline quality with reduced scattering from deficiencies of our samples.

\begin{figure*}
  \centering
  \includegraphics[width=6in]{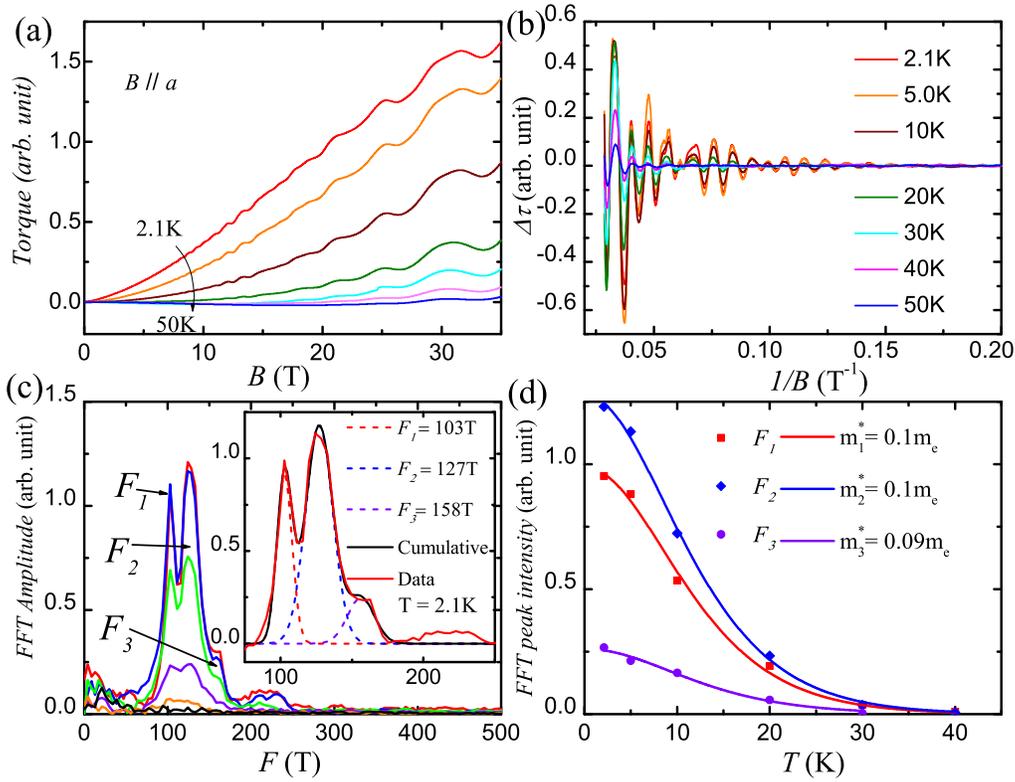}
  \caption{(a) Field dependence of magnetic torque $\tau$ for SrZnSb$_2$ at different temperatures with magnetic field nearly parallel with out-of-plane direction. (b) The oscillatory part of the magnetic torque $\Delta\tau$. (c) The FFT spectra for $\Delta\tau$ between 5 T and 15 T at different temperatures. Three major oscillation frequencies are indicated. Inset: multi-peak Gaussian fit of the FFT spectra to obtain the intensity of the three frequencies (only 2 K data is shown). (d) The temperature dependence of the FFT amplitude for the three frequencies and the fits to thermal damping term of LK formula (solid lines).
  }
  \label{fig:Fig2}
\end{figure*}

We have performed magnetic torque measurements up to 35 T and probed strong dHvA oscillations in SrZnSb$_{2}$, as seen in Fig. 2(a). Figure 2(b) shows the dHvA oscillations after a polynomial background subtraction. The oscillations start from 5 T at $T =$ 2.1 K and remain up to 50 K in high field range. As seen in Fig. 2(a), the oscillation peak at 30 T tends to broaden as the temperature goes down, implying that the Zeeman effect takes place in the high field range. To avoid any possible influence by Zeeman effect, we have performed the fast Fourier transform (FFT) in the field range between 5 T and 15 T as shown in Fig. 2(c), and resolved three frequencies with $F_{1} =$ 103 T, $F_{2} =$ 127 T, and $F_{3} =$ 160 T seen in the inset of Fig. 2(c). The effective masses of the quasiparticles corresponding to each of the oscillation frequencies can be obtained by fitting the temperature dependence of the FFT peak amplitude to the thermal damping term ${R_T}$ in  Lifshitz-Kosevich (LK) formula\cite{LK_formula}, i.e. $\frac{{\Delta \rho }}{\rho } \propto \alpha T\mu /\left[ {\bar {B}sinh\left( {\alpha T\mu /\bar {B}} \right)} \right]$, where $1/\bar {B}$ is the average inverse field when performing FFT analysis. As shown in Fig. 2(d), the fits result in three effective masses of 0.1 m$_{e}$, 0.1 m$_{e}$ and 0.09 m$_{e}$ for $F_{1}$, $F_{2}$ and $F_{3}$, respectively. The quasiparticle effective masses in SrZnSb$_{2}$ are slightly lighter than those found in the Bi based 112 compounds like BaZnBi$_{2}$ \cite{Wang_BaZnBi2, Ren_BaZnBi2, Zhao_BaZnBi2}, but heavier than those in other Sb based 112 compounds \cite{Liu_SrMnSb2, Liu_BaMnSb2}.

\begin{figure*}
  \centering
  \includegraphics[width=6in]{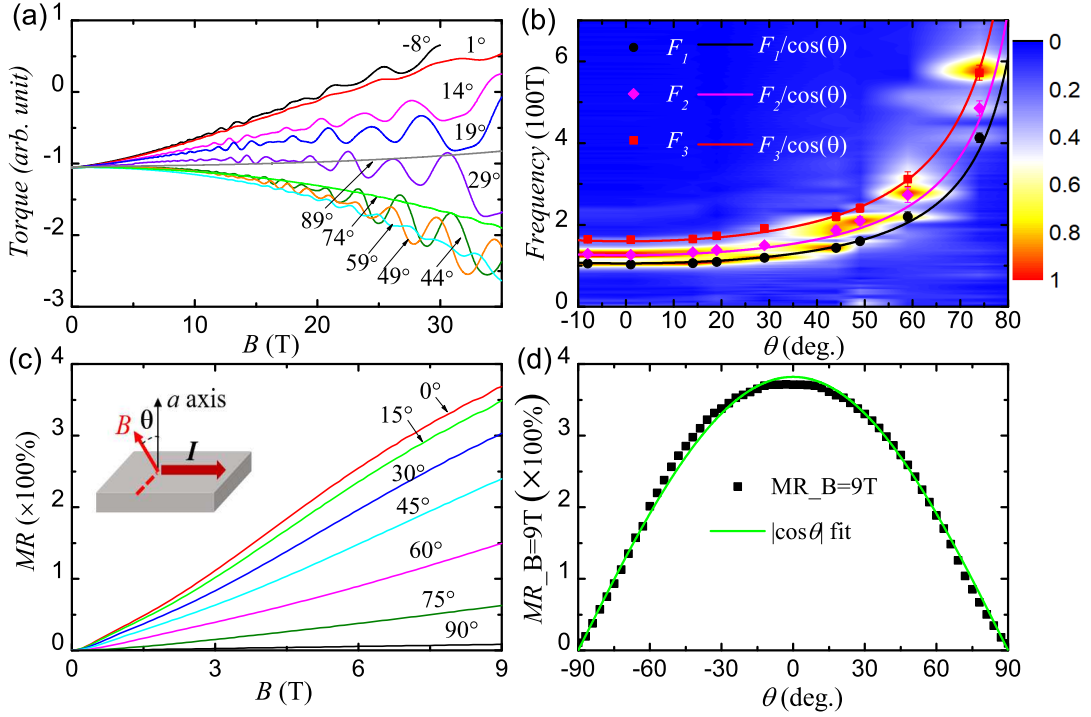}
  \caption{(a) Field dependence of the magnetic torque measured with different field orientations. (b) The angular dependence of the dHvA oscillation frequency $F (\theta)$ and the fits to $F(\theta ) = {F_0}/\left| {\cos \theta } \right|$. The background is the 2D color plot for angle dependent FFT spectra of the dHvA oscillations. (c) Magnetoresistance of SrZnSb$_2$ measured with different field orientations. Inset: A schematic diagram of the measurement geometry. (d) The angular dependence of resistivity at $B= 9 T$ and the $\left| {\cos \theta } \right|$ fit.
  }
  \label{fig:Fig4}
\end{figure*}

To map out the Fermiology of SrZnSb$_{2}$, the evolution of the magnetic torque properties with field orientation has been investigated. As seen in Fig. 3(a), the dHvA oscillations show systematic shifts to higher fields as the field is tilted away from $a$ axis where the angle is defined as $\theta = {0^ \circ}$. Figure 3(b) presents a color plot of the angular dependent FFT spectra, from which the revolution of FFT peaks can be clearly tracked. The obtained oscillating frequencies are plotted against the angle $\theta$. $F_{1}$, $F_{2}$ and $F_{3}$ are well fitted by $F(\theta )={F_0}/\left| {\cos \theta } \right|$, which suggests the corresponding Fermi surfaces corresponding to the quantum oscillations are of quasi-2D like. Figure 3(c) shows the $MR (= [{\rho _{xx}}(B) - {\rho _{xx}}(0)]/{\rho _{xx}}(0))$ data measured at 2 K up to 9 T with different angles between the magnetic field and sample $bc$ plane. The field is kept perpendicular to the in-plane current direction during rotation. When the field is applied perpendicularly to the sample $bc$ plane, MR is $\sim$ 370\% at 9 T. It gradually becomes smaller as the field is tilted away and almost vanishes when the field is parallel with the $bc$ plane. Figure 3(d) shows the MR measured with the field rotating about the current direction at $B =$ 9 T and $T =$ 2 K. The result can be well described by $MR \propto |\cos \theta|$, i.e. the MR is only related to the field component perpendicular to the $bc$ plane. This has been seen in other 112 material, like Ca/SrMnBi$_2$ \cite{Wang_CaMnBi2, Wang_SrMnBi2}, which confirms the quasi-2D nature of the transport.

To investigate if SrZnSb$_2$ has non-trivial topology, Shubnikov-de Haas (SdH) oscillations have been studied. The in-plane magnetotransport properties with a magnetic field perpendicular to the plane up to 35 T are presented in Fig. 4. $\rho_{xx}$ exhibits linear field dependence at low field ($B<$ 10 T), and tends to saturate at high fields with a magnetoresistance (MR) of $\sim 700$\% at 35 T. $\rho_{xy}$ is positive and linearly dependent with field for $B<$ 5 T, and then becomes flat at higher fields, indicating multiple band transport nature with the dominant carriers being holes. The change of slope near 10 T implies the variation of densities or mobilities for electron- or hole-like carriers at high field range.

\begin{figure*}
  \centering
  \includegraphics[width=7.0in]{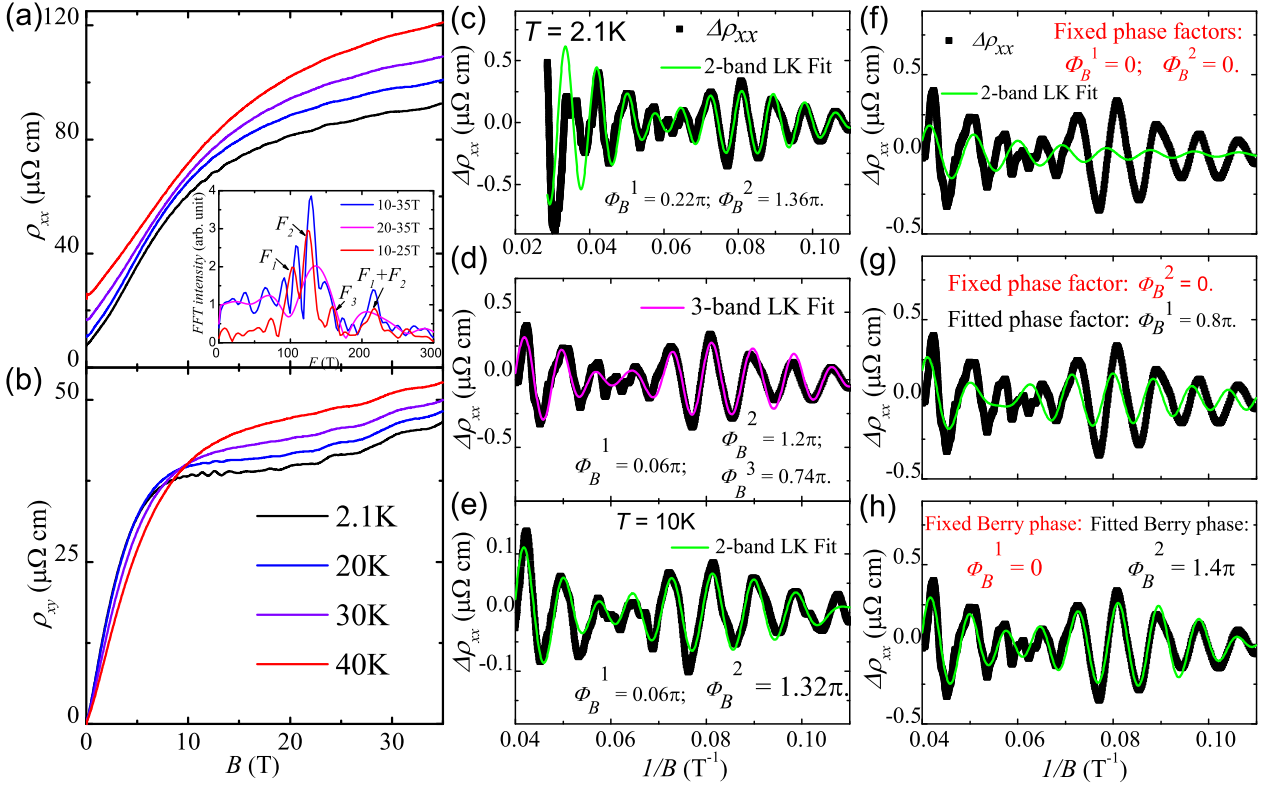}
  \caption{(a) Field dependence of in-plane resistivity $\rho_{xx}$ of SrZnSb$_2$ measured up to 35 T at different temperatures. Inset: The comparison of the FFT spectra made between 10 T and 35 T, 20 T to 35 T and 10 T to 25 T at 2 K. (b) Hall resistivity $\rho_{xy}$ up to 35 T measured simultaneously with $\rho_{xx}$. (c) The oscillatory part of resistivity $\triangle\rho_{xx}$ (black dots) at $T =$ 2.1 K and the fit to LK formula with two known frequencies, namely $F_{1}$ and $F_{2}$. (d) The fit of SdH oscillations at $T =$ 2.1 K to LK formula including three frequencies, which are $F_{1}$, $F_{2}$ and $F_{3}$. (e) shows the two-band LK fits of the SdH oscillations at higher temperatures, $T =$ 10 K, resulting similar Berry phases to those of SdH oscillations at $T =$ 2.1 K. (f), (j) and (h) are the trials to fit the SdH oscillations at $T =$ 2.1 K to two-band LK formula with fixed Berry phase(s) for $F_1$ and/or $F_{2}$. (f) The Berry phases are both fixed to be 0 for $F_{1}$ and $F_{2}$. (j) Only the Berry phase of $F_{2}$ is fixed to be 0, and it's a free fitting parameter for $F_{1}$. (h) Only the Berry phase of $F_{1}$ is fixed to be 0, and it's a free fitting parameter for $F_{2}$.
  }
  \label{fig:Fig4}
\end{figure*}

As shown in Figs. 4(a) and (b), SdH oscillations are observable for both $\rho_{xx}$ and $\rho_{xy}$ when $B > 10$ T. Since Zeeman effect can be observed at high fields, FFT from 10 T to 25 T for the $\rho_{xx}$ at $T =$ 2.1 K is made, resulting in the dominant oscillating frequencies of 103 T and 127 T with a third minor frequency of 160 T (see the inset of Fig. 4(a)), consistent with the values we obtained in dHvA. The SdH oscillations can be described by the LK formula \cite{LK_formula}: $$\Delta \rho  \propto {B^\lambda }{R_T}{R_D}{R_S}\cos [2\pi (\frac{F}{B} + \gamma  - \delta )]$$ where the thermal damping term ${R_T}$ is equal to $\alpha T\mu /\left[ {Bsinh\left( {\alpha T\mu /B} \right)} \right]$ and the Dingle damping factor ${R_D} = exp\left( { - \alpha {T_D}\mu /B} \right)$. $\mu$ is the ratio of cyclotron effective mass ${m^*}$ to the free electron mass ${m_e}$ and $\alpha  = \left( {2{\pi ^2}{k_B}{m_e}} \right)/(\hbar e)$. ${T_D}$ is the Dingle temperature. $R_S = \cos (\pi g \mu/2)$ is the spin damping factor, where $g$ is the Landau $g$ factor. $\lambda$ is a dimensional factor, which is 1/2 for the 3D case and 0 for the 2D case. The phase factor $\gamma-\delta$ of the cosine term is linked with the Berry phase $\Phi _B$ by the relation $\gamma  = \frac{1}{2} - \frac{{{\Phi _B}}}{{2\pi }}$. \cite{Berry_phase} The phase shift $\delta$ is 0 for 2D Fermi surfaces and $ \pm 1/8$ for 3D ones depending on the FS cross-section and the carrier type \cite{Berry_phase 3D}.

\begin{figure*}
  \centering
  \includegraphics[width=7.0in]{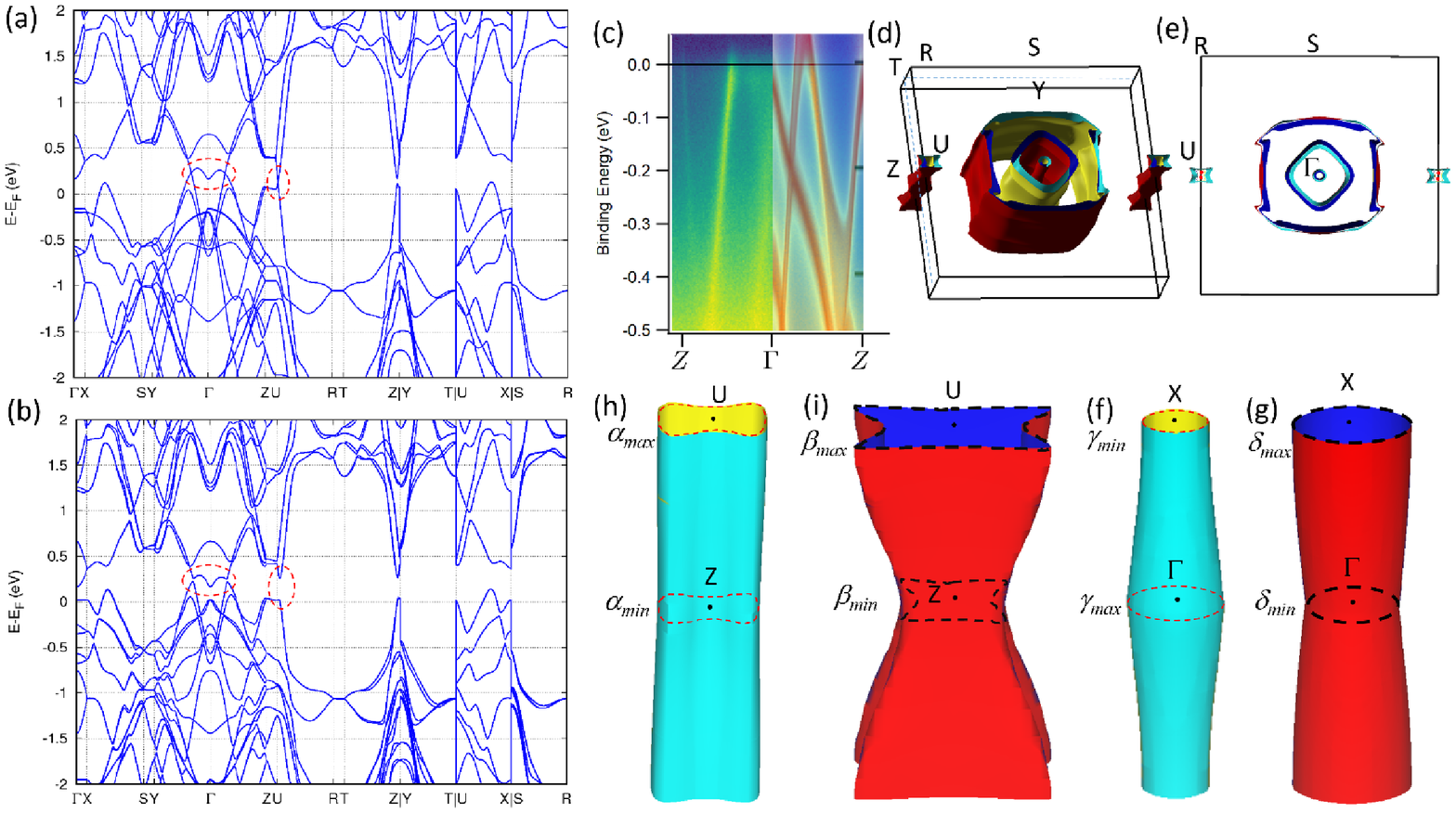}
  \caption{(a)-(b) Band structure of SrZnSb$_2$ without SOC (a) and with SOC (b). The Dirac nodes are circled by red dash lines. The Fermi level has been adjusted based on the ARPES data. (c) The energy band spectra along $Z-\Gamma-Z$ line observed by ARPES and its fit with the calculations. (d)-(e) Fermi Surface of SrZnSb$_2$ from the side (d) and top view (e). (h)-(g) Individual $\alpha$, $\beta$, $\gamma$ and $\delta$ Fermi pockets. The maximum and minimum cross section areas of each pocket are circled by the dash lines.
  }
  \label{fig:Fig5}
\end{figure*}

To examine the Berry phase accumulated from the cyclotron motions of the quasiparticles in the system, as the FFT spectra are dominated by $F_{1}$ and $F_{2}$, we first fit the SdH oscillations ($\Delta \rho_{xx}$) at $T =$ 2.1 K to the generalized LK formula with two frequencies \cite{Hu_TaP}, which we will refer to as the two-band LK formula. As seen in Fig. 4(c), the data can be fit well to the two-band LK formula for $1/B>0.04 $ T$^{-1}$. If the higher field ($B>25$ T) data is included, the fit cannot converge. This discrepancy the data and the LK formula fit at high fields probably come from the Zeeman splitting of the Landau levels. Besides, from the FFT with different field ranges we note that the FFT spectra peaks tend to move to higher values when the the higher field range (B $> 25$ T) is included. This gives us a hint the Zeeman effect appears at high fields. Similar effect has previously been observed in WTe$_2$, where the oscillatory data can not be reproduced by LK formula for the entire field range because of the strong Zeeman effect at high fields \cite{WTe2_Zeeman}. As such, we focus our fitting at the field range below 25 T where the influence from Zeeman effect should be minimized as evidenced by the high fitting quality. The fitted phase factors $\gamma-\delta$ are 0.39 and 0.82. Considering the 2D Fermi surface revealed from the angular dependence of the oscillations which will be shown later, the $\delta$ is taken to be 0. Thus, the obtained Berry phases are 0.22$\pi$ and 1.36$\pi$ for $F_1$ and $F_2$, respectively. The former indicates a trivial Berry phase for $F_1$, while the latter represents a nontrivial Berry phase for $F_2$. We have also tried to include the third minor frequency $F_3$ for the fitting. The SdH data is fitted by three-band LK formula (Fig. 4(d)). The obtained Berry phase for $F_3$ is $0.74 \pi$, close to a nontrivial value $\pi$. The resulting Berry phases for $F_1$ and $F_2$ are 0.06 $\pi$ and 1.2 $\pi$, respectively, which are consistent with those from two-band LK fit in Fig. 4(c). As the three-band LK formula fit does not show notable improvement of the fitting quality, in order to minimize the amount of fitting parameters, we believe the two-band LK fit should be enough to capture the oscillatory amplitudes and phases of the data.

To check the reliability of the phase factors obtained by the two-band LK fit, we attempted to fit the SdH oscillations at $T =$ 2.1 K to the LK formula using fixed phase factors (Fig. 4 (f-h)). In Fig. 4(f), the phase factors $\gamma-\delta$ for both frequencies in two-band LK formula are fixed to 0.5 so that the Berry phases are both trivial. However, as shown clearly in Fig. 2 (f), the data can not be fitted at all in this case. In Fig. 4(g), the Berry phase for $F_2$ is fixed to 0 and the phase factor of $F_1$ is a free fitting parameter. The best fit gives a significant phase offset between the raw data and the fitting line. Only if the Berry phase for $F_1$ is fixed to be 0 and the phase factor of $F_2$ is a free fitting parameter, the data can be well fitted and the obtained Berry phase for $F_2$ is close to the original two-band LK fit where no phase factor is fixed as seen in Fig. 4(h). Furthermore, to demonstrate the stability of the two-band LK fitting, the same phase factors obtained in Fig. 4(c) are used in the fitting of the SdH oscillations at $T =$ 10 K. As shown in Figs. 4(e), these phases factors fit the high temperature data well. From the LK fit, we have clearly demonstrated the trivial Berry phase for $F_1$ and a nontrivial case for $F_2$ and $F_3$, suggesting the existence of Dirac fermions.

To better understand the electronic structure of SrZnSb$_2$, we have performed Density functional theory (DFT) calculations with and without SOC and discovered that SrZnSb$_2$ is a nodal-line Dirac semimetal. As shown in Fig. 5(a), there are three Dirac points in the energy band diagram when switching off SOC, which agree well with the results from Ref. \cite{TQ_chemistry, TM_prediction}. The two Dirac points in $\Gamma-Z$ line and $\Gamma-Y$ line are the cross-sectional points of a nodal line lying in $Y-Z$ plane. This node-ring is protected by symmetry operation $\{m_{100}|1/2,1/2,1/2\}$ because the two bands, which cross to form this node-ring, have opposite eigenvalues of operator $\{m_{100}|1/2,1/2,1/2\}$ in $Y-Z$ plane. In the presence of SOC, this nodal line is gapped because two spin components in the same energy band have opposite eigenvalues of $\{m_{100}|1/2,1/2,1/2\}$. The energy gap opened at the nodal line is small, about 30 meV, comparable to 20 meV of the SOC induced gap at the Dirac line nodes in ZrSiS \cite{ZrSiS_Node-line_ARPES}. As shown in Fig. 5 (c), the DFT calculation agrees well with the ARPES spectra, showing a strong, dispersive electron-like band crossing at $E_F$ and another electron-like band at the $Z$ point. Based on our calculated band structure and ARPES spectra, the Dirac node is about 200 meV above the Fermi level of our sample.

The Dirac point lying in $U-R$ line is a double Dirac point formed by the crossing of two doubly degenerate bands without SOC, which is protected by nonsymmorphic symmetry. Following the notion of Ref. \cite{Aroyo_notion}, the eigenstates of the two crossing bands belong to the representation of $P_1+P_3$ and $P_2+P_4$, respectively, leading to crossing of the two bands rather than anticrossing. When SOC is included, we find the bands along $U-R$ remain four-fold degeneracy considering spin component. However, there is only one irreducible representation along $U-R$ line in this situation. As such, the crossing of each pair of bands is forbidden and the Dirac point in this line is gapped accordingly. The energy gap is about 230 meV. In addition, we note that this Dirac point is quasi-2D since both its conduction and valence bands have nearly flat dispersions along $Z-U$ direction.

We determine the Fermi level for the sample by adjusting the DFT energy bands with those from ARPES along $Z-\Gamma-Z$ line, as shown in Fig. 5(c). We  obtain an experimental Fermi energy $E_F$ to be 267 meV lower than the theoretical value, consistent with the observation that the transport properties are dominated by holes. This shifted $E_F$ may be caused by possible Sr/Zn deficiencies in the sample. Based on the adjusted $E_F$, we plot the Fermi surface in Fig. 5 (d) and (e). One should keep in mind that this Fermi energy cuts two valence bands, resulting in two sets of Fermi surface of similar shape with one set nested into the other.

All of Fermi surfaces are 2D like with slight corrugation along $k_x$ direction, consistent with the 2D nature of the Fermiology revealed by our dHvA data. The three oscillation frequencies observed in our quantum oscillations can only come from the two small non-trivial Fermi pockets $\alpha$ and $\beta$ centering at the BZ edge Y point and the other two small trivial Fermi pockets centering at the BZ center $\Gamma$ point, as shown in Figs. (h)-(g). We have calculated the frequencies associated with the extreme area of these four Fermi pockets using the code provided by ref. \cite{FS_cal} and summarized them in the Table III.

\begin{table}
\caption{The comparison of oscillation frequencies obtained from DFT calculations and experimental observations with $B // a$ axis of SrZnSb$_2$. To avoid the effect of Zeeman effect on the data analysis, FFT was done on the SdH data (5 T to 15 T) and dHvA data (10 T to 25 T). Based on the DFT calculations, $\alpha$ and $\beta$ Fermi pockets have non-trivial topology while $\gamma$ and $\delta$ Fermi pockets are trivial. }
\begin{tabular}{l c c c c c c c c}
\hline
\hline

& $\alpha_{min} $  & $\alpha_{max} $ & $\beta_{min} $  & $\beta_{max} $& $\gamma_{min} $  & $\gamma_{max} $& $\delta_{min} $  & $\delta_{max} $     \\
\hline
F$^{\rm DFT}$ (T) & 109 & 125 & 153 & 368 & 47 & 106 & 115 & 183 \\
F$^{\rm dHvA}$ (T) & - & 127 & 160 & - & - & 103 & - &-\\
F$^{\rm SdH}$ (T) & - & 127  & 160  & - & - & 103  & - &-\\
m$^{\rm dHvA}$ (m$_e$) & - & 0.1 & 0.09 & - & - & 0.1 & - &-\\
$\Phi^{\rm SdH}$   & - & 1.2$\pi$  & 0.74$\pi$  & - & - & 0.06$\pi$  & - &-\\

\hline

\end{tabular}
\label{tab.3}
\end{table}

\section{Discussion}

Here we would like to have more discussions on the Berry phase for the three frequencies observed from QO. In theory, $F_{1}$ from a trivial fermi surface should correspond to a Berry phase ($\phi_B$) of 0, which is consistent with our experimental value $\phi_{B}^{\alpha} = 0.06\pi$. And each of the Fermi surfaces that are responsible for $F_{2}$ and $F_{3}$ includes two equivalent Dirac points. As is mensioned above, these Dirac points are quasi-2D, so every Dirac point should contribute a Berry phase of $\pi$ similar to the case of graphene. Two Dirac points will lead to a Berry phase of 2$\pi$ for both $F_{2}$ and $F_{3}$. However, the gaps opened at the two Dirac points will alter the Berry phase according to the equation $\phi_B = \pm 2 \pi (1-\Delta/2E_{FC})$, where $\Delta$ is the gap size and $E_{FC}$ represents energy difference between fermi energy and the center of energy gap\cite{Berry_phase 3D}. The Berry phases for $F_{2}$ and $F_{3}$ are estimated to be 1.57$\pi$and 0.36$\pi$, respectively. This explains the Berry phases we obtained from quantum oscillations are not close to $\pi$ for $F_{2}$ and $F_{3}$, while it clearly shows that the Fermi surfaces that are responsible for the two frequencies are topological nontrivial.

Since the nodal-line and nodal features are about 200 meV above the FL in our SrZnSb$_2$ sample, our current ARPES experiments only observe linear dispersive bands but not the nodal features, however, our comprehensive study of QO, DFT and ARPES data can unambiguously prove the existence of the non-trivial topology. In the future study, the sample will be electron doped by substituting La into the Sr sites to tune the FL above the nodal features and then ARPES will be performed to investigate the emergent phenomena associated with the topological nodal-line, such as the drumhead states, etc.

Finally, to understand the role of magnetism in the AMnX$_2$ materials, we compare the relativistic fermion transport properties between SrZnSb$_{2}$ and its magnetic version Sr$_{1-y}$Mn$_{1-z}$Sb$_{2}$ \cite{Liu_SrMnSb2}. The two compounds have the same orthorhombic crystal structure with space group of $Pnma$, and the Sb square nets are distorted in both compounds. In Sr$_{1-y}$Mn$_{1-z}$Sb$_{2}$, the canted AFM order of Mn moments coexists with the topological state hosted by Sb distorted square net layers. As a result, the relativistic transport properties, including the amplitude and frequency of the SdH oscillations, are coupled with the FM component. In both materials, the Fermi surfaces consist of relativistic fermions are quasi-2D like. However, because of the lack of AFM layers between two neighboring distorted Sb square net layers, the resistivity anisotropy of SrZnSb$_{2}$ is much weaker. Further doping studies of SrMn$_{1-x}$Zn$_{x}$Sb$_{2}$ would be beneficial for understanding the effect of magnetism on Dirac fermion transport in the system.

\section{Conclusions}
In conclusion, we have performed a comprehensive study of the quantum transport properties and the electronic structure of SrZnSb$_2$ with a combination of the magnetotransport measurements, first-principles calculations and ARPES experiment. From the analysis of quantum oscillations, low effective masses and non-trivial Berry phases are revealed. These quantum oscillations are revealed to be associated with the small trivial Fermi pockets in the BZ center and the nontrivial Fermi pockets in the BZ edge. In addition, a node line surrounding the BZ center with tiny SOC gap is suggested by the first-principles calculations.

\section*{Acknowledgments}
Work at UCLA was supported by DOE Award DE-SC0011978. The National High Magnetic Field Laboratory is supported by the National Science Foundation
through NSF/DMR-1644779 and the State of Florida. Work at ORNL was sponsored by the Scientific User Facilities Division, Office of Basic Energy Sciences, DOE. HBC acknowledges the support from Office of Science, Office of Basic Energy Sciences, DOE under Award Number DE-AC05-00OR22725. P.L. and Q.L. acknoledge the support from the Guangdong Innovative and Entrepreneurial Research Team Program under Grant No. 2017ZT07C062 and Guangdong Provincial Key Laboratory for Computational Science and Material Design under Grant No. 2019B030301001. Work at the University of Colorado was supported by the U.S. Department of Energy (DOE), Office of Science, Office of Basic Energy Sciences under Award  DE-FG02-03ER46066. The ARPES experiment used resources of the Advanced Light Source, which is a DOE Office of Science User Facility under contract no. DE-AC02-05CH11231.

\end{document}